\input{aipcheck}

\documentclass[
    ,final            
  ]
  {aipproc}

\layoutstyle{6x9}

\begin{document}

\title{Non-empirical nuclear energy functionals, pairing gaps and odd-even mass differences}

\classification{21.60.Jz, 21.10.Dr}
\keywords      {Non-empirical energy density functional, finite nuclei, odd-even mass staggering}
\author{T.~Duguet}
{address={CEA, Centre de Saclay, IRFU/Service de Physique Nucl\'eaire, F-91191 Gif-sur-Yvette, France},altaddress={National Superconducting Cyclotron Laboratory and Department of Physics and Astronomy, Michigan State University, East Lansing, MI 48824, USA}}

\author{T.~Lesinski}
{address={Department of Physics and Astronomy, University of Tennessee, Knoxville, TN 37996, USA},altaddress={Physics Division, Oak Ridge National Laboratory, Oak Ridge, TN 37831, USA}}

\begin{abstract}
First, we briefly outline some aspects of the starting project to design non-empirical energy functionals based on low-momentum vacuum interactions and many-body perturbation theory. Second, we present results obtained within an approximation of such a scheme where the pairing part of the energy density functional is constructed at first order in the nuclear plus Coulomb two-body interaction. We discuss in detail the physics of the odd-even mass staggering and the necessity to compute actual odd-even mass differences to analyze it meaningfully.
\end{abstract}

\maketitle


\section{Introduction and elements of formalism}
\label{intro}

Like-particle pairing is an essential ingredient of nuclear-structure models, in particular regarding the description of exotic nuclei~\cite{doba03a}. Also, superfluidity plays a key role in neutron stars, e.g. it impacts post-glitch timing observations~\cite{Avogadro07} or their cooling history~\cite{heiselberg3}.

Within a single-reference (SR) implementation of the energy density functional (EDF) formalism~\cite{bender03a}, pairing is incorporated through the breaking of the $U(1)$ symmetry associated with particle-number conservation. As a result, the binding energy ${\cal E}_{SR}$ of the many-body system is postulated to be a functional of both the one-body density matrix $\rho_{ji} \equiv \langle\Phi| c^\dagger_i c_j |\Phi\rangle$ and the pairing tensor $\kappa_{ji} \equiv \langle\Phi| c_i c_j |\Phi\rangle$, the dependence on the latter being allowed by the use of an auxiliary product state of reference $|\Phi\rangle$ that mixes particle numbers (of given parity)~\cite{RingSchuck}.

Modern empirical parameterizations of existing EDFs, e.g. Skyrme or Gogny, provide a fair description of bulk and certain spectroscopic properties of known nuclei~\cite{bender03a}. On the other hand, they lack predictive power away from known data and a true spectroscopic quality, in particular regarding the part that drives superfluidity. As a result, several groups currently work on empirically improving the analytical form and the fitting of functionals, e.g. see Refs.~\cite{margueron08a,yamagami09a} for recent attempts to pin down the isovector content of purely local pairing functionals.

Along with improving the phenomenology at play, the quest for predictive EDFs starts to benefit from a complementary approach~\cite{Duguet06} that does not primarily rely on fitting known data but that roots the analytical form of the functional and the value of its couplings into underlying low-momentum two- and three-nucleon (NN and NNN) interactions~\cite{Bogner03,roth08a} through the application of many-body perturbation theory\footnote{Infinite resummation of certain categories of diagrams and/or a redefinition of the unperturbed vacuum $|\Phi\rangle$ are always possible. Switching from conventional hard-core potentials to low-momentum interactions is essential to make a perturbative approach viable, e.g. second-order calculations performed in terms of low-momentum interactions provide satisfactory results for bulk correlations~\cite{Bogner05,Bogner09a,roth06a}.} (MBPT)~\cite{Nozieres}. The overall goal of such a project is (i) to bridge with {\it ab-initio} many-body techniques applicable to light nuclei, (ii) calculate properties of heavy/complex nuclei from basic vacuum interactions and (iii) perform controlled calculations with theoretical error bars. First results following such a route are currently being reported~\cite{drut09a}. It is an objective of the present contribution to expose results of such an effort to build the pairing part of the EDF non-empirically~\cite{Duguet04,Duguet07,Lesinski09a,KaiAchimTD,Lesinski08}.

We propose to write the energy functional at a given order in (Goldstone) MBPT under a generic form that is convenient to bridge with existing phenomenological EDFs
\begin{eqnarray}
\label{eq:e00}
{\cal E}_{SR}[\{\rho_{ij}\}, \{\kappa_{ij}\}, \{\kappa^{\ast}_{ij}\}; \{E_k\}]
& \equiv & \sum_{ij} t_{ij} \, \rho_{ji} \\
&&  + \frac{1}{2} \sum_{ijkl} \bar{v}^{\rho\rho}_{ijkl} \,
        \rho_{ki} \, \rho_{lj}
      + \frac{1}{4} \sum_{ijkl} \bar{v}^{\kappa\kappa}_{ijkl} \,
        \kappa^{\ast}_{ij} \, \kappa_{kl} \nonumber \\
&& +  \frac{1}{6} \sum_{ijklmn} \bar{v}^{\rho\rho\rho}_{ijklmn} \,
        \rho_{li} \, \rho_{mj} \, \rho_{nk}
+ \frac{1}{4} \sum_{ijklmn} \bar{v}^{\rho\kappa\kappa}_{ijklmn} \, \rho_{li} \,
        \kappa^{\ast}_{jk} \, \kappa_{mn} \nonumber \\
&& + \frac{1}{24} \sum_{ijklmn} \bar{v}^{\rho\rho\rho\rho}_{ijklmnop} \,
        \rho_{mi} \, \rho_{nj} \, \rho_{ok} \, \rho_{pl} + \ldots  , \nonumber
\end{eqnarray}
where all dependencies on $\rho$ and $\kappa^{\ast}\kappa$ have been made explicit. The {\it effective vertices} $\bar{v}^{\rho\rho}_{ijkl}$, $\bar{v}^{\kappa\kappa}_{ijkl}$\ldots thus introduced are expressed in terms of the vacuum two-, three-,\dots body  interactions and on quasi-particle energies $E_k$ that are to be determined self-consistently through a chosen procedure. More precisely, a term of given power in $\rho$ and/or $\kappa^{\ast}\kappa$ in Eq.~\ref{eq:e00} receives contributions from different perturbative orders and/or many-body forces. To exemplify this, we can write the vertices arising at second order in the NN interaction $\bar{v}^{NN}$, in a perturbation theory that does not account for pairing explicitly\footnote{The single-particle basis solution of Eq.~\ref{hfb} also diagonalizes the density matrix $\rho$ of $|\Phi\rangle$ in this case.}
 \begin{eqnarray}
\label{eq:e01}
\bar{v}^{\rho\rho\rho\rho}_{ijklijkl} \equiv  6 \,
\frac{\bigl|\bar{v}^{NN}_{ijkl}\bigr|^2}{\epsilon_{i} + \epsilon_{j} - \epsilon_{k} - \epsilon_{l}}  \, \,  \, \, \, ;  \, \, \, \, \, \bar{v}^{\rho\rho\rho}_{ijkijk} \equiv \frac{1}{2} \sum_{l} \bar{v}^{\rho\rho\rho\rho}_{ijklijkl}  \, \, \, \, \, ;  \, \, \, \, \,  \bar{v}^{\rho\rho}_{ijij} \equiv \bar{v}^{NN}_{ijij} + \frac{1}{6} \sum_{k} \bar{v}^{\rho\rho\rho}_{ijkijk} \, \, \, ,
\end{eqnarray}
where $\epsilon_{i}$ denotes single-particle energies to be determined self-consistently. The EDF form of Eq.~\ref{eq:e00} may naively suggests that it results from the average value, in the unperturbed vacuum, of an (hypothetical) effective Hamilton operator containing two-body (second line), three-body (third line),\dots pieces. However, Eq.~\ref{eq:e01}, that provides microscopic expressions for the matrix elements of $\bar{v}^{\rho\rho}_{ijkl}$, $\bar{v}^{\kappa\kappa}_{ijkl}$, $\bar{v}^{\rho\rho\rho}_{ijklmn}$\ldots, demonstrates that re-extracting an (effective) Hamilton operator from the energy density has no foundation\footnote{Note for instance that symmetry properties of $\bar{v}^{\rho\rho}_{ijkl}$, $\bar{v}^{\rho\rho\rho}_{ijklmn}$ and $\bar{v}^{\rho\rho\rho\rho}_{ijklijkl}$ under the exchange of fermionic indices are {\it not} as expected from two-, three- and four-body operators.} and can at best be the result of approximations.

Forms as given by Eq.~\ref{eq:e00} are known as orbital-dependent energy functionals~\cite{engel03a} in electronic systems density functional theory (DFT), with the important subtlety that DFT implies that quasi-particle wave functions $(U_k,V_k)$ and quasi-particle energies $E_k$ are generated through the variationally optimum local one-body potential, i.e. the optimal effective potential (OEP)~\cite{talman76a}. We do not insist on that here to rely on a framework that embraces empirical Gogny functionals whose associated one-body fields are non-local~\footnote{Using a non-local pairing field, as in the present work, renormalizes from the outset the ultraviolet divergence that arises when using a (quasi-)local pairing field.}. Of course, none of the existing empirical functionals do depend on quasi-particle energies $E_k$. It remains to be seen in the future whether such an extension is necessary and tractable.

An alternative to OEP that is closer to what is currently done with empirical EDFs consists of determining quasi-particle wave-functions and energies through the minimization of ${\cal E}_{SR}$ with respect\footnote{Not only the present scheme does not insist on obtaining a local potential but also partial derivatives with respect to quasi-particle energies are omitted. The so-called Krieger-Li-Iafrate approximation to the OPE~\cite{kli90a} also omits such functional derivatives.} to independent matrix elements of $\rho$ and $\kappa$, under the constraint to have given neutron and proton numbers in average. This leads to solving Hartree-Fock-Bogoliubov-like (HFB)~\cite{RingSchuck} equations
    \begin{eqnarray}
    \label{hfb}
     \left(
     \begin{array}{cc}
      h - \lambda  & \Delta \\
      -\Delta^\ast & -h^\ast + \lambda \\
     \end{array}
     \right)
     \left(
     \begin{array}{c}
      U_k \\
      V_k \\
     \end{array}
     \right)
     &=& E_k
     \left(
     \begin{array}{c}
      U_k \\
      V_k \\
     \end{array}
     \right) \, \, \, .
    \end{eqnarray}

The one-body field $h$ that drives the correlated single-particle motion and the shell structure, as well as the field $\Delta$ that drives superfluidity, are defined as
    \begin{equation}
     h_{ij} \equiv \frac{\delta {\cal E}_{SR}}{\delta \rho_{ji}} \equiv t_{ij}+\Sigma_{ij}  \equiv t_{ij}+\sum_{kl} \overline{v}^{ph}_{ikjl} \; \rho_{lk} \hspace{0.5cm} ;  \hspace{0.5cm}
     \Delta_{ij} \equiv \frac{\delta {\cal E}_{SR}}{\delta \kappa^\ast_{ij}} \equiv \frac{1}{2}
        \sum_{kl} \overline{v}^{pp}_{ij kl}\; \kappa_{kl} \, \, ,
    \end{equation}
through which two effective vertices $\overline{v}^{ph}$ and $\overline{v}^{pp}$ are introduced that can be expressed in terms of $\bar{v}^{\rho\rho}_{ijkl}$, $\bar{v}^{\kappa\kappa}_{ijkl}$\ldots, i.e. they themselves possess a diagrammatic expansion in terms of $\bar{v}^{NN}$ and $\bar{v}^{NNN}$.

Our immediate focus is on the pairing part of the EDF. Beyond enhancing its predictive power, our aim is to understand better the microscopic processes that build superfluidity in finite nuclei. Typical questions relate to (i) the contribution from the direct NN and NNN interactions, its breaking down in partial waves (essentially $^1S_0$, $^3P_1$, $^1D_2$ in decreasing order of expected importance), as well as (ii) the role of higher-order effects associated with the coupling to (collective) fluctuations.

To answer the first of these two questions, our current target is to perform reliable finite-nuclei calculations at first order in low-momentum NN and NNN interactions generated through renormalization group techniques~\cite{Bogner03,roth08a}. The upper row of Tab.~\ref{tab:exp_schemes} shows the corresponding diagrammatic for the one-body fields, omitting for simplicity contributions from the NNN interaction. As an intermediate step, we present here approximate results such that $\overline{v}^{ph}$ and the part of the EDF that depends only on $\rho$ is empirically provided by the SLy4 Skyrme parametrization~\cite{chabanat98}, while pairing vertices $\bar{v}^{\kappa\kappa}=\overline{v}^{pp}$ are computed at first order in the Coulomb plus nuclear\footnote{We use the low-momentum NN interaction V$_{{\rm low \, k}}$~\cite{Bogner03} built from the Argonne $v_{18}$ NN potential~\cite{Wiringa95} at a renormalization cut-off $\Lambda=2.5$\,fm$^{-1}$.} NN interaction~\cite{Duguet04,Duguet07,Lesinski09a,KaiAchimTD}. Such an EDF contains neither energy dependencies nor $\rho$-$\kappa$ cross terms. Only the dominant $^{1}S_0$ partial-wave of the NN is included in $\bar{v}^{\kappa\kappa}$ whereas the effect of $^{3}P_1$ and $^{1}D_2$ is discussed in Ref.~\cite{baroni09a}. Also, the first-order contribution of the NNN interaction to $\bar{v}^{\rho\kappa\kappa}$ and $\overline{v}^{pp}$ will be reported on in Ref.~\cite{Lesinski08}. Note that for such a calculation to be a decent approximation of the targeted first-order one, at least as for extracting pairing gaps, it is crucial that the empirical Skyrme parametrization that drives the underlying shell structure is characterized by an isoscalar effective k-mass $m^{\ast}_0\approx 0.7\;m$ at saturation density~\cite{KaiAchimTD}. Of course, we eventually aim at calculating $h$ at lowest-order in {\it both} the NN and the NNN low-momentum interactions, possibly making use of the density matrix expansion~\cite{bogner08a,gebremariam09a}. Eventually, higher-order contributions are left out for future works\footnote{See the lower row of Tab.~\ref{tab:exp_schemes} for the second-order contributions to the one-body fields.}.

\begin{table}
\vspace{2pt}
\begin{tabular}{lll|lll}
\parbox[l][1.0cm][c]{0.8cm}{$\Sigma^{(1)}=$} &\parbox[l][1.0cm][c]{0.8cm}{\includegraphics[width=0.8cm]{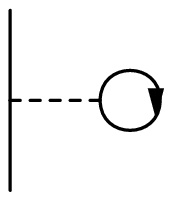}} & & &
\parbox[l][1.0cm][c]{1.0cm}{$\Delta^{(1)}=$} &\parbox[l][1.0cm][c]{0.8cm}{\includegraphics[width=0.8cm]{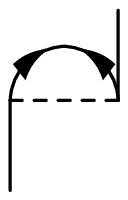}} \\ \hline\hline
\parbox[l][1.0cm][c]{0.8cm}{$\Sigma^{(2)}=$} &
\parbox[l][1.0cm][c]{0.8cm}{\includegraphics[width=0.8cm]{plots/SHF_1.eps}} \parbox[l][1.0cm][c]{0.35cm}{$\:+$} \parbox[l][1.0cm][c]{0.8cm}{\includegraphics[width=0.8cm]{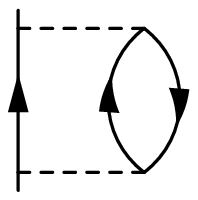}} \parbox[r][1.0cm][c]{0.35cm}{$\:+$} \parbox[l][1.0cm][c]{0.8cm}{\includegraphics[width=0.8cm]{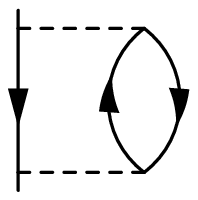}} & & &
\parbox[l][1.0cm][c]{0.8cm}{$\Delta^{(2)}=$} &
\parbox[l][1.0cm][c]{0.65cm}{\includegraphics[width=0.8cm]{plots/SHF_2.eps}} \parbox[l][1.0cm][c]{0.35cm}{$\:+$} \parbox[l][1.0cm][c]{0.8cm}{\includegraphics[width=0.8cm]{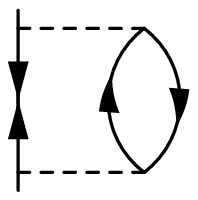}}
\end{tabular}
\caption{Perturbative expansion scheme to first (up) and second (down) order. The dashed line denotes the free-space NN interaction. Diagrams with more than one anomalous propagator are not shown.}
\label{tab:exp_schemes}
\end{table}

\section{Experimental versus theoretical pairing gaps}

We limit ourselves to discussing the odd-even mass staggering (OEMS) whereas other observables are reported on in Ref.~\cite{Lesinski08}. The OEMS is dominated by the deficit of binding energy of the unpaired nucleon in odd nuclei, i.e. the "pairing gap". In the SR-EDF formalism, such a staggering relates to the description of odd nuclei through the excitation of a quasi-particle on top of an even-number parity vacuum. Experimentally, the OEMS is extracted through $n$-points difference-mass formulae $\Delta^{(n)}_{q}(N/Z)$~\cite{bohr69a}. However, because of the technical difficulty to compute odd nuclei, data are often compared to purely theoretical pairing gaps extracted from the calculation of just one even-even nucleus. One such theoretical pairing gap is the {\it Lowest Canonical State} gap $\Delta_{\rm LCS}$, defined as the diagonal matrix element of the pairing field $\Delta$ in the \emph{canonical} single-particle state whose associated quasi-particle energy $E_k$ is the lowest~\cite{Duguet07,Lesinski09a}.

The difficulty with such comparisons is not only that (i) finite-difference mass formulae are contaminated by contributions other than the targeted "pairing gap"~\cite{satula98a,rutz99a,bender00a,Duguet01a,Duguet01b} but also that (ii) the "pairing gap" that makes the actual OEMS is itself an average of $\Delta_{\rm LCS}$ extracted from the even-even and (blocked) odd-even nuclei involved in the finite-difference mass formula~\cite{Duguet01a,Duguet01b}. As a result, comparisons based on theoretical gaps extracted from one even-even nucleus can only be of semi-quantitative character, which is often fine as empirical pairing functionals are not yet targeting a nucleus by nucleus agreement with experiment. However, aiming at such a level of agreement in the (distant?) future and at doing so in a non-empirical fashion requires the comparison of apples with apples, i.e. to compare theoretical and experimental odd-even mass differences. At the price of requiring a good understanding of the different contributions to the OEMS~\cite{Duguet01a,Duguet01b}, doing so allows more fruitful comparisons between theory and experiment, e.g. to analyze the interplay between pairing and the underlying shell structure. This is what we wish to briefly exemplify in the present contribution.

\section{Results}

\begin{figure}
  \includegraphics[height=.35\textheight]{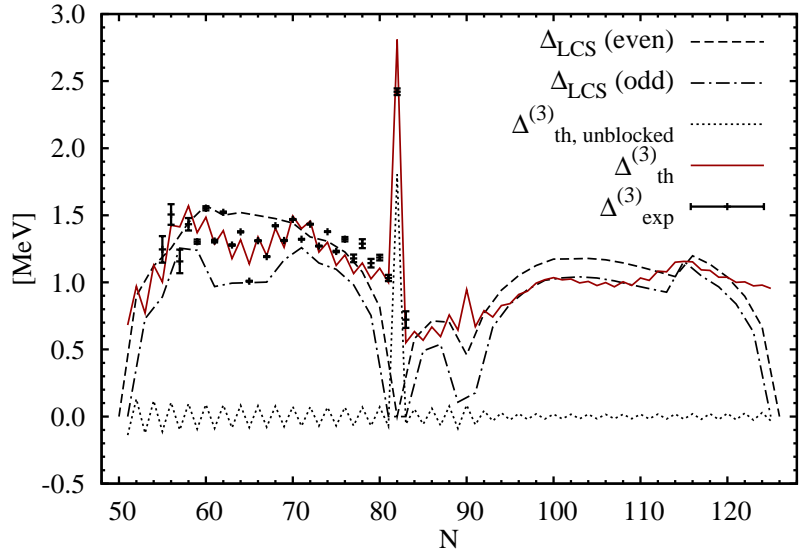}
  \caption{Experimental three-point mass differences (crosses) for neutrons along the tin isotopic chain versus several theoretical measures of the OEMS: $\Delta^n_{\rm LCS}({\rm even})$ (dashed-line), $\Delta^n_{\rm LCS}({\rm odd})$ (dashed-dotted line) and $\Delta^{(3)}_{n}(N)$ (full line). Theoretical three-point mass differences are also shown for odd-even nuclei computed using an even-number-parity vacuum as a reference state, i.e. a HFB state without {\it any} quasi-particle blocking as if odd-even nuclei had the same structure as even-even ones (dotted line)~\cite{Duguet01a,Duguet01b}.}
  \label{OEMStin}
\end{figure}

In Refs.~\cite{Duguet07,Lesinski09a}, experimental $\Delta^{(3)}_{q}({\rm odd})$ were compared to $\Delta^q_{\rm LCS}({\rm even})$. Main results were that neutron and proton pairing gaps computed from the Skyrme plus non-empirical pairing energy functional were close to data for a large set of semi-magic light-, medium- and heavy-mass nuclei. Implications of such results were also discussed. Here, we wish to analyze the qualitative modifications brought about by comparing directly theoretical and experimental three-point mass differences. To do so, we computed odd-even nuclei through the self-consistent blocking procedure performed within the filling approximation~\cite{perezmartin08a,bertsch09a}. Results for neutron gaps along the tin isotopic chain are reported in Fig~\ref{OEMStin}.

To analyze meaningfully the OEMS~\cite{Duguet01b}, the ground state of odd nuclei is best understood as a quasi-particle excitation on top of an even-number parity vacuum that shares the structure of even-even nuclei but that has the odd number of particles on average. In this way, the quasi-particle excitation is performed at (almost) constant particle number~\cite{Duguet01a}. The even-number parity vacuum provides the smooth part of the energy while the quasi-particle excitation, that is dominated by the static "pairing gap", generates the relative lack of binding of odd nuclei without which no interesting physics would be extracted from odd-even mass differences. The dotted line in Fig.~\ref{OEMStin} shows the contribution of the smooth part of the energy to $\Delta^{(3)}_{n}(N)$, i.e. when odd-even isotopes are described as if they had the structure of even-even ones. One sees that such a contribution, which reflects the curvature of the smooth part of the energy, oscillates {\it symmetrically around zero} and accounts exactly for the odd-even oscillation of $\Delta^{(3)}_{n}(N)$.
This demonstrates that contributions other than the targeted "pairing gap" contaminate $\Delta^{(3)}_{n}(N)$ {\it in an opposite way} for odd and even $N$~\cite{Duguet01b}, which contradicts the usual belief~\cite{satula98a} that $\Delta^{(3)}_{n}({\rm odd})$ is free from such contaminations\footnote{It was suggested in Ref.~\cite{Duguet01b} to use $\Delta^{(3)}_{n}({\rm odd})$ as a measure of the sole "pairing gap" {\it because} the contribution from the time-odd reversal symmetry breaking, not discussed in the present paper, possibly cancels out the contribution from the smooth part of the energy in this case. This is however subject to revision due to the current lack of knowledge regarding time-odd terms in the nuclear EDF.}.

In Fig.~\ref{OEMStin}, the comparison between experimental data for odd $N$ and $\Delta^q_{\rm LCS}({\rm even})$ (dashed line) recalls the results of Refs.~\cite{Duguet07,Lesinski09a} and sets the stage for what comes next. Those two curves are consistent with each other, with a slight overestimation (underestimation) of the data at mid-shell (just below and above the $N=82$ shell closure). Such a situation is representative of the results obtained along other semi-magic isotopic and isotonic chains. Still, certain features that are visible in the data, i.e. (i) the lowering around $N=65$, (ii) the flat trend as one approaches the $N=82$ shell closure and (iii) the finite jump from $N=81$ to $N=83$, are not reproduced by the bell-shaped curve provided by $\Delta^q_{\rm LCS}({\rm even})$. At best, one can talk of an overall semi-quantitative agreement and wonder whether the remaining discrepancies are due to limitations of (a) the pairing part of the EDF, (b) the use of $\Delta^q_{\rm LCS}({\rm even})$ as a measure of the OEMS and/or (c) SR calculations that miss dynamical correlations associated with particle number restoration and collective pairing vibrations that are of importance in the weak pairing regime, e.g. near shell closures.

The full-fledged comparison of experimental (stars) and theoretical (full line) three-point mass differences is also provided in Fig.~\ref{OEMStin}. The most striking feature is the ability of the calculation to grasp quantitatively the three non-trivial features seen in the data and outlined in the previous paragraph. As a result, one goes from a semi-quantitative agreement with experiment across the major shell using $\Delta^q_{\rm LCS}({\rm even})$ to the ability to compare on a nucleus by nucleus basis. In particular, there were hints that the lowering of the pairing gaps around $N=65$ could be partly due to dynamical pairing fluctuations~\cite{anguiano02a}. Here, such a feature is well reproduced at the SR-HFB level. It will be of interest to study whether using the non-empirical pairing functional computed from the finite-range and non-local V$_{{\rm low \, k}}$ interaction is essential to obtain such a pattern or if it is entirely driven by the interplay with the underlying shell structure, independently of the detailed characteristics of the pairing functional employed.

A similar situation occurs regarding the behavior of the OEMS towards and across the $N=82$ shell closure. One sees from $\Delta^q_{\rm LCS}$ that static pairing correlations collapse in the immediate vicinity of $N=82$, i.e. in $^{131,132,133}$Sn, while the experimental $\Delta^{(3)}_{n}(N)$ sustains a non-zero value down to $N=81$ and $83$\footnote{One must remove $\Delta^{(3)}_{n}(82)$ from the analysis as it measures the $N=82$ shell gap rather than static pairing correlations. Contrarily, $\Delta^{(3)}_{n}(81,83)$ are {\it not} influenced by the $N=82$ shell gap.}. Surprisingly enough, when going from $\Delta^q_{\rm LCS}$ to theoretical three-point mass differences, the experimental trend is well captured down to $N=81$ and across the $N=82$ where the OEMS jumps by $350$ keV, $\Delta^{(3)}_{n}(83)$ being the last piece of available data. Due to the collapse of $\Delta^q_{\rm LCS}$ close to the shell closure, it is usually stated that $\Delta^{(3)}_{n}$ is dominated by other contributions than static pairing in this regime, i.e. by dynamical pairing fluctuations and contributions associated with the discreteness of the underlying shell structure. Regarding the former, we just saw that a SR calculation omitting entirely dynamical pairing fluctuations\footnote{The Lipkin-Nogami procedure is not used in the present calculation.} can account for the data. Regarding the latter, it is to be noted that (i) in the (hypothetical) zero-pairing limit, and still assuming spherical symmetry, $\Delta^{(3)}_{n}(N)$ is zero from $N=70$ to $N=81$ as one fills the highly degenerate $h_{11/2}$ shell and that (ii) the regularly oscillating contribution of the smooth part of the energy seen in Fig.~\ref{OEMStin} demonstrates that the structure of odd nuclei is still best understood, down to $^{131}$Sn and $^{133}$Sn, as a quasi-particle excitation on top of a {\it statically paired} even-number parity vacuum. Eventually, the energy of the quasi-particle excitation that builds $\Delta^{(3)}_{n}(81,83)$ and leads to the unpaired {\it blocked} state is dominated by pairing correlations, i.e. it would be zero in the zero-pairing limit. Although dynamical pairing correlations are likely to renormalize the OEMS, the present results implies that odd-even mass differences might be less impacted by such correlations than other observables in the vicinity of shell closures.

\section{Conclusions}

 We discuss pairing gaps obtained in tin isotopes using an energy density functional whose pairing part is constructed at first order in the nuclear plus Coulomb interaction. Only the (dominant) $^{1}S_0$ partial wave of the two-nucleon force is incorporated whereas the contributions from $^{3}P_1$ and $^{1}D_2$~\cite{baroni09a}, as well as from the three-nucleon interaction~\cite{Lesinski08}, will be reported on soon. Most importantly, we discuss in detail the physics of the odd-even mass staggering and the necessity to compute actual odd-even mass differences to analyze it meaningfully and compare with data on a nucleus-by-nucleus basis. In particular, an excellent description of the odd-even mass staggering is obtained in the vicinity of magic shell closures {\it prior} to incorporating dynamical pairing correlations associated with particle number restoration and pairing vibrations.


\begin{theacknowledgments}
We wish to thank K. Bennaceur, K. Hebeler, J. Meyer and A. Schwenk for our fruitful collaboration on designing non-empirical pairing energy density functionals. This work was supported by the U.S. Department of Energy under Contract Nos. DE-FG02-96ER40963, DE-FG02-07ER41529 (University of Tennessee) and DE-AC05-00OR22725 with UT-Battelle, LLC (Oak Ridge National Laboratory).

\end{theacknowledgments}



\bibliographystyle{aipproc}   


\begin{thebibliography}{9}

\bibitem{doba03a}
J. Dobaczewski, W. Nazarewicz, Prog. Theor. Phys. Suppl. \textbf{146} (2003) 70.

\bibitem{Avogadro07}
P. Avogadro, F. Barranco, R. A. Broglia, E. Vigezzi, Phys. Rev. C \textbf{75} (2007) 012805.

\bibitem{heiselberg3}
H. Heiselberg, M. Hjorth-Jensen, Phys. Rep. \textbf{328} (2000) 237.

\bibitem{bender03a}
M. Bender, P.-H. Heenen, P.-G. Reinhard, Rev. Mod. Phys. \textbf{75} (2003) 121.

\bibitem{RingSchuck}
P. Ring, P. Schuck,  \emph{The Nuclear Many-Body Problem}, Springer, Berlin, Heidelberg, (2000).

\bibitem{margueron08a}
J. Margueron, H. Sagawa, K. Hagino , Phys. Rev. C \textbf{77} (2008) 054309

\bibitem{yamagami09a}
M. Yamagami, Y. R. Shimizu, T. Nakatsukasa, arXiv:0812.3197

\bibitem{Duguet06}
T. Duguet, K. Bennaceur, T. Lesinski, J. Meyer,
\emph{Opportunities with Exotic Beams}, Proceedings of the
3rd ANL/MSU/JINA/INT RIA Workshop, edited by T.
Duguet, H. Esbensen, K. M. Nollett, C.D. Roberts (World
Scientific, 2007) p. 21; nucl-th/0606037.

\bibitem{Bogner03}
S. K. Bogner, T. T. S. Kuo, A. Schwenk, Phys. Rep. \textbf{386} (2003) 1.

\bibitem{roth08a}
R. Roth, S. Reinhardt, H. Hergert, Phys. Rev. C \textbf{77} (2008) 064003

\bibitem{Bogner05}
S. K. Bogner, A. Schwenk, R. J. Furnstahl, A. Nogga, Nucl. Phys. A \textbf{763} (2005) 59.

\bibitem{Bogner09a}
S. K. Bogner, R. J. Furnstahl, A. Nogga, A. Schwenk, arXiv:0903.3366.

\bibitem{roth06a}
R. Roth, P. Papakonstantinou, N. Paar, H. Hergert, T. Neff, H. Feldmeier, Phys. Rev. C \textbf{73} (2006) 044312.

\bibitem{Nozieres}
P. Nozi\`eres, \emph{Theory of interacting Fermi systems}, Westview press, Advanced Book Classics, (1964).

\bibitem{drut09a}
J. E. Drut, R. J. Furnstahl, L. Platter, arXiv:0906.1463, and references therein

\bibitem{Duguet04}
T. Duguet, Phys. Rev. C \textbf{69} (2004) 054317.

\bibitem{Duguet07}
T. Duguet, T. Lesinski, Eur. Phys. Jour. ST \textbf{156} (2008) 207.

\bibitem{Lesinski09a}
T. Lesinski, T. Duguet, K. Bennaceur, J. Meyer, Eur. Phys. J. A \textbf{40} (2009) 121.

\bibitem{KaiAchimTD}
K. Hebeler, T. Duguet, T. Lesinski, A. Schwenk,  arXiv:0904.3152

\bibitem{Lesinski08}
T. Lesinski, T. Duguet, K. Bennaceur, J. Meyer, in preparation.

\bibitem{engel03a}
E. Engel, \emph{A Primer in Density Functional Theory}, edited by C. Fiolhais, F. Nogueira and M. Marques,
(Springer, Berlin, 2003), p. 56.

\bibitem{talman76a}
J. D. Talman, W. F. Shadwick, Phys. Rev. A \textbf{14} (1976) 36

\bibitem{kli90a}
J. B. Krieger, Y. Li, G. J. Iafrate, Phys. Lett. A \textbf{146} (1990) 256

\bibitem{chabanat98}
E. Chabanat, P. Bonche, P. Haensel, J. Meyer, R. Schaeffer, Nucl. Phys. A \textbf{635} (1998) 231.

\bibitem{Wiringa95}
R. B. Wiringa, V. G. J. Stoks, R. Schiavilla, Phys. Rev. C \textbf{51} (1995)  38.

\bibitem{baroni09a}
S. Baroni, A. Schwenk, in preparation

\bibitem{bogner08a}
S. K. Bogner, R. J. Furnstahl, L. Platter, arXiv:0811.4198

\bibitem{gebremariam09a}
B. Gebremariam, S. K. Bogner, T. Duguet, in preparation

\bibitem{bohr69a}
A. Bohr and B. R. Mottelson, \emph{Nuclear Structure} (Benjamin,
New York, 1969), Vol. 1

\bibitem{satula98a}
W. Satula, J. Dobaczewski, W. Nazarewicz, Phys. Rev. Lett. \textbf{81} (1998) 3599.

\bibitem{rutz99a}
K. Rutz, M. Bender, P.-G. Reinhard, J. A. Maruhn, Phys. Lett. B \textbf{468} (1999) 1.

\bibitem{bender00a}
M. Bender,  K. Rutz, P.-G. Reinhard, J. A. Maruhn, Eur. Phys. J. A \textbf{8} (2000) 59.

\bibitem{Duguet01a}
T. Duguet, P. Bonche, P.-H. Heenen, J. Meyer, Phys. Rev. C \textbf{65} (2001) 014310.

\bibitem{Duguet01b}
T. Duguet, P. Bonche, P.-H. Heenen, J. Meyer, Phys. Rev. C \textbf{65} (2001) 014311.

\bibitem{perezmartin08a}
S. Perez-Martin, L. M. Robledo, Phys. Rev. C \textbf{78} (2008) 014304.

\bibitem{bertsch09a}
G. F. Bertsch, C. A. Bertulani, W. Nazarewicz, N. Schunck, M. V. Stoitsov, Phys. Rev. C \textbf{79} (2009) 034306.

\bibitem{anguiano02a}
M. Anguiano, J. L. Egido, L. M. Robledo, Phys. Lett. B \textbf{545} (2002) 62.

\end{thebibliography}


\end{document}